\documentclass{INTERSPEECH2023}

\usepackage{amsmath,graphicx,amssymb, cite,multirow,booktabs,bm,changepage}
\usepackage{ragged2e} 
\usepackage[linesnumbered,ruled,vlined]{algorithm2e}
\usepackage{algpseudocode}
\usepackage{amsmath}

\interspeechcameraready

\title{Few-shot Class-incremental Audio Classification Using Adaptively-refined Prototypes}

\name{Wei Xie$^1$,Yanxiong Li$^1$, Qianhua He$^1$, Wenchang Cao$^1$, Tuomas Virtanen$^2$ \thanks{Corresponding author : Yanxiong Li. This work was supported by the national natural science foundation of China (62111530145, 61771200), Guangdong basic and applied basic research foundation (2022A1515011687, 2021A1515011454), and international scientific research collaboration project of Guangdong (2021A0505030003). }}

\address{\normalsize{$^1$School of Electronic and Information Engineering, South China University of Technology, Guangzhou 510641, China\\
}$^2$Audio Research Group, Tampere University, Tampere, Finland}

\email{chester.w.xie@gmail.com, eeyxli@scut.edu.cn}
\begin{document}

\maketitle
 
\begin{abstract}
New classes of sounds constantly emerge with a few samples, making it challenging for models to adapt to dynamic acoustic environments. This challenge motivates us to address the new problem of few-shot class-incremental audio classification. This study aims to enable a model to continuously recognize new classes of sounds with a few training samples of new classes while remembering the learned ones. To this end, we propose a method to generate discriminative prototypes and use them to expand the model's classifier for recognizing sounds of new and learned classes. The model is first trained with a random episodic training strategy, and then its backbone is used to generate the prototypes. A dynamic relation projection module refines the prototypes to enhance their discriminability. Results on two datasets (derived from the corpora of Nsynth and FSD-MIX-CLIPS) show that the proposed method exceeds three state-of-the-art methods in average accuracy and performance dropping rate.

\end{abstract}
\noindent\textbf{Index Terms}: few-shot learning, class-incremental learning, meta-training, audio classification

\section{Introduction}

\label{sec:intro}

Audio classification is a basic task for many audio-related applications, such as acoustic scene classification \cite{xie2021acoustic,xie2022deep,hasan2022genetic}, sound event detection \cite{mesaros2021sound,imoto2022impact,park2022cross}, and bioacoustic monitoring \cite{cramer2020chirping,teixeira2022fledge,gatto2023discriminative}. Deep learning has been successfully applied to various audio processing tasks in recent years \cite{cosbey2019deep,morrone2021audio}. However, these successes mainly rely on large amounts of labeled data, and tedious model tuning \cite{zhao2022self,chen2022end}. Moreover, the models adopted in these works are generally trained in a supervised manner, which makes the models only recognize classes defined at the development stage \cite{teh2021open}. If some classes (e.g., rare animal sounds) are not included in the development dataset, the deployed model cannot recognize those audio classes \cite{tavares2022open}.

To improve the ability of the model to adapt to dynamic acoustic environments, some researchers have introduced the few-shot learning (FSL) paradigm to the audio domain \cite{huang2021unsupervised,wang2022hybrid,huang2022meta}. However, these FSL-based studies mainly focus on improving the model's ability to recognize new audio classes but do not pay much attention to maintaining the model's ability to remember learned ones. Humans can recognize new audio classes with few samples or even one sample without forgetting the learned ones. Such human capacity is needed to improve the intelligence of the model. Therefore, we introduce the few-shot class-incremental learning (FSCIL) paradigm \cite{tao2020few, zhu2021self} to the audio domain and address a new problem of few-shot class-incremental audio classification (FCAC).

The problem setup of FCAC is different from the existing FSL-based audio processing tasks \cite{morfi2021few,yang2022mutual,Liu2022a}. Specifically, the former required the model to continually recognize new classes of sounds without forgetting the learned ones. In contrast, the latter only required the model to recognize new classes of sounds. Recently, Wang et al. \cite{wang2021few} proposed a few-shot continual learning framework for audio classification using the dynamic few-shot learning (DFSL) method \cite{gidaris2018dynamic}. They expand the classifier of the model with an attention-based weights generator (ABWG) to recognize both new and old sound classes. Their framework assumes a scenario where the model needs to recognize both old and new classes within a single session. However, the FCAC problem requires the model to perform recognition across multiple continuous sessions and distinguish between previously learned classes and newly encountered ones. Thus, the FCAC problem poses greater challenges and represents real-world scenarios of intelligent audio processing.


Most deep models can be decoupled into a backbone and a classifier \cite{zhang2021few}. Therefore, to expand a model for classifying new and learned audio classes, the classifier of the model needs to be expanded with new classification weights or prototypes \cite{snell2017prototypical}. To that end, one can expand the classifier with randomly initialized prototypes for new audio classes and then retrain the model using the methods such as the finetune method \cite{zhou2021pycil} and the method of incremental classifier and representation learning (ICaRL) \cite{rebuffi2017icarl}. Alternatively, we extend the solution proposed by Wang et al. \cite{wang2021few} for the FCAC problem by reusing the ABWG to expand the classifier in each session. Therefore, we provide three baseline methods for the FCAC problem. 

\begin{figure*}[ht]
  \centering
  \includegraphics[scale=0.61]{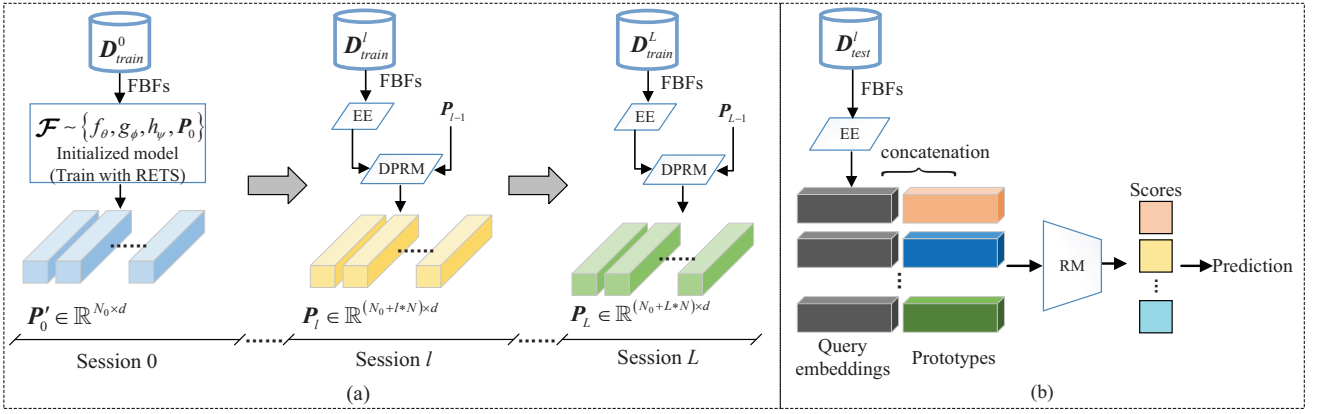}
  \caption{Schematic diagram of the proposed method for FCAC. (a) The process of prototype generation and refinement. (b) Classification of the test samples using the RM and the prototypes.}
  \label{fig_proposed_scheme}
  \vspace{-1.5em}
\end{figure*}

To tackle the FCAC problem, we propose a method to generate discriminative prototypes and use them to expand the model's classifier. The proposed method consists of the random episodic training strategy (RETS) and the dynamic relation projection module (DRPM). The RETS is designed to train the initial model with strong generalization, and the DRPM is designed to refine the prototypes to make the model more discriminative. Two audio datasets, namely Neural synthesis of 100 classes of musical instrument sound (NSynth-100) and Freesound clips of 89 classes (FSC-89), are developed for performance evaluation of all methods. Experimental results on the two datasets show that the proposed method outperforms the baseline methods in average accuracy (AA) and performance dropping rate (PD). The contributions of this work are mainly two-fold. First, we address the new problem of FCAC, which aims to makes the model incrementally recognize new classes with few training samples without forgetting learned classes. Second, we provide baseline solutions and propose an alternative method for tackling the FCAC problem.

\section{Problem Description}
\label{sec:problem_description}

The FCAC problem involves $L$ learning sessions that come in sequence. The first session is called the base session, and the remaining ones are called incremental sessions. A model for FCAC is developed in the base session and evaluated in all incremental sessions. Let $\bm{D}_{train}^{l}=\left \{ \left ( x^{l}_{i} ,y^{l}_{i}  \right )  \right \}_{i=1}^{n_{train}^{l} } $ and $\bm{D}_{test}^{l}=\left \{ \left ( \hat{x}^{l}_{j} ,\hat{y}^{l}_{j}  \right )  \right \}_{j=1}^{n_{test}^{l} }$ denote the training set and the test set in session $l\left ( 0\le l\le L \right ) $, respectively. $x^{l}_{i}$ and $\hat{x}^{l}_{j}$ denote the training sample and the test sample of session $l$, respectively. $y^{l}_{i}$ and $\hat{y}^{l}_{j}$ are the label of $x^{l}_{i}$ and $\hat{x}^{l}_{j}$, respectively. $n_{train}^{l}$ and $n_{test}^{l}$ are the number of samples in $\bm{D}_{train}^{l}$ and $\bm{D}_{test}^{l}$, respectively. $\bm{D}_{train}^{l}$ and $\bm{D}_{test}^{l}$ have the same label set $\bm{Y}_{l}$, i.e., $y^{l}_{i},\hat{y}^{l}_{j}\in \bm{Y}_{l}$. The label sets of any two training sets are disjoint, i.e., $\forall l\ne k,\bm{Y}_{l} \cap \bm{Y}_{k}= \varnothing$. Among all $L$ training sets, only $\bm{D}_{train}^{0}$ contains a substantial amount of training samples. In contrast, each of the remaining training sets contains only a few training samples, which are organized in the format of $N$-way $K$-shot, i.e., there are $N$ classes of data in the training set, and each class has $K$ training samples. Due to data privacy and the shortage of memory, the training data in prior sessions (sessions 0 to $l$-1) are generally discarded after usage and thus only $\bm{D}_{train}^{l}$ can be used to update the model $\bm{\mathcal{F}}$ in session $l$. Meanwhile, the updated model needs to make predictions on test sets of both the current session and all prior sessions, i.e., $ \bm{D}_{test}^{0}\cup  \bm{D}_{test}^{1}\cup \cdots \cup \bm{D}_{test}^{l} $. As a result, the objective for solving the FCAC problem is to minimize the empirical risk \cite{zhou2022forward} of the model over all test sets:
\begin{equation}
  \min \left \{  {\textstyle \sum_{\left ( \hat{x}_{j},\hat{y}_{j}   \right )\in D_{test}^{0} \cup \dots \cup D_{test}^{L}  }^{}} \mathcal{L} \left ( \bm{\mathcal{F}}\left ( \hat{x}_{j}  \right ) ,\hat{y}_{j}  \right )  \right \} 
\end{equation} 
where $\mathcal{L}\left ( \cdot ,\cdot  \right ) $ denotes a loss function.

\section{Method}
\label{sec:method}
The motivation of the proposed method is to expand the classifier of the model with discriminative prototypes. Each prototype \cite{snell2017prototypical} is initialized as the mean of all the embeddings of training samples from that class:
\begin{equation}
  \bm{p}_{k} =\frac{1}{\left | \bm{S}_{k}  \right | } \sum_{\left ( x_{i} ,y_{i}  \right )\in \bm{S}_{k}}^{ } f_{\theta } \left ( x_{i}  \right )  
\end{equation} 
where $\bm{p}_{k}$ and $\bm{S}_{k} $ are the prototype and the training set of class $k$, respectively. $f_{\theta } $ denotes an embedding extractor (EE).

An EE with strong generalization ability can be more robust across different audio classes, which results in more semantically meaningful embeddings and prototypes that capture the essence of different audio classes. In addition, the generalization ability of the EE is critical to avoid the impact of noise and redundant information on the prototypes. Therefore, one way to obtain discriminative prototypes is to improve the generalization ability of the EE. To this end, we propose using RETS to train the EE by simulating the test scenario, where the training data is organized to mimic the evaluation process. Besides, due to the limited number of training samples used to derive prototypes of new classes, their discriminability may be weaker than that of the audio classes in the base session. This can result in an unbalanced discriminability between the new and learned prototypes, leading to degraded classification performance of the model. To address this challenge, we propose using DRPM to refine the prototypes of new and learned classes to minimize their confusion. The refined prototypes can be more well-separated, thus enhancing the model's classification performance.

As shown in Fig. \ref{fig_proposed_scheme} (a), the number of prototypes expands incrementally session by session. In session $0$, a model consisting of four modules is trained using the RETS with $\bm{D}_{train}^{0}$. The first module is an EE $f_{\theta } :\mathbb{R} ^{D} \longrightarrow \mathbb{R} ^{d}$, which is used to convert the filter-bank features (FBFs) of audio samples into embeddings. $D$ and $d$ denote the dimension of FBFs and embeddings, respectively. The second module is a DRPM $g_{\phi }$, which is used to refine the prototypes. The third module is a relation module (RM) $h_{\psi}$ for implementing the classification. The fourth module is the initial prototypes of base classes, which is represented by a learnable parameter matrix $\bm{P}_{0} \in \mathbb{R}^{N_{0}\times d } $. $N_{0}$ is the number of base classes.

In each incremental session, the EE takes the training set of the new classes and generates prototypes for those classes. Then, the DPRM is utilized to refine the new prototypes with the prototypes of the previous session. After that, predictions of the test samples can be made based on the similarity between the embedding of each test sample and the prototypes of all classes in that session. As is shown in Fig. \ref{fig_proposed_scheme} (b), a relation module (RM) is used as a learnable metric \cite{sung2018learning} to calculate the similarity between the prototypes and the test embedding. Besides, to prevent the learned knowledge from being forgotten in incremental sessions, the parameters of EE, DRPM, and RM obtained in the base session are frozen in all incremental sessions. Instead, the prototypes are continually expanded and refined to recognize new and learned audio classes.

\vspace{-1.0em}
\begin{algorithm}[ht]

  \KwIn{\justifying \noindent $\bm{D}_{train}^{0}$; A randomly initialized model $\bm{\mathcal{F} }$ including four modules: $f_{\theta }$, $g_{\phi }$, $h_{\psi}$ and  $\bm{P}_{0}$. The number of episode in each training iteration $t$.}

  \KwOut{\justifying \noindent A trained model.}

  \While{\rm{not convergence}}{
    \justifying \noindent Randomly sample a test subset $\bm{Q}$ from $\bm{D}_{train}^{0}$, whose label set is $\bm{Y}_{Q}$, and $\bm{Y}_{Q} \subset  \bm{Y}_{0} $\;

    \justifying \noindent Randomly sample $t$ sets of $N$-way $K$-shot training subsets from $\bm{D}_{train}^{0}$: $\left \{ \bm{S}_{1},\cdots, \bm{S}_{t}   \right \}$, whose label sets are $\left \{ \bm{Y}_{{\bm{S}}_{1}},\cdots, \bm{Y}_{{\bm{S}}_{t}}   \right \}$ \;
    
    \justifying \noindent \For{i \rm{in} $[1,t]$}{
    \justifying \noindent Keep the prototypes in $\bm{P}_{0}$, whose labels are not in $\bm{Y}_{{\bm{S}}_{i}}$ and treat them as the pseudo-base prototypes $\bm{P}_{0}{}'$\;

    \justifying \noindent Use $f_{\theta }$ and ${\bm{S}}_{i}$\ to generate the prototypes of pseudo-new classes $\bm{P}_{\bm{S}_i}$;

    \justifying \noindent Merge $\bm{P}_{0}{}'$ and $\bm{P}_{S_i}$ to form the initial prototypes in an episode $\bm{P}_{init}$\;

    \justifying \noindent Take $\bm{P}_{init}$ into DRPM to obtain the refined prototypes: $\bm{P}_{re}= g_{\phi }(\bm{P}_{init},\bm{P}_{0}{}')$

    \justifying \noindent Obtain predictions $\bm{\hat{Y}} _{Q}$ of the test embeddings $f_{\theta }\left ( \bm{Q} \right )$ using $\bm{P}_{re}$ and the RM\;

    \justifying \noindent Calculate and accumulate the loss: $\sum \mathcal{L}_{CE} \left ( \hat{\bm{Y}} _{Q}, \bm{Y} _{Q} \right ) $ \;
  
  }
  \justifying \noindent Optimize the model with SGD algorithm\;
  }
  \caption{ Random episodic training strategy. $\mathcal{L}_{CE} \left ( \cdot , \cdot  \right )$ denotes the cross-entropy loss function. }\label{algo_RETS}
  
\end{algorithm}
\vspace{-1.5em} 

\subsection{Random episodic training strategy}
\label{ssec:RETS}
The random episodic training strategy (RETS) is to organize the training data into a series of simulated tasks and use them to optimize the model. Those tasks mimic the few-shot circumstances encountered in each incremental session. The pseudo-code of the RETS is presented in Algorithm \ref{algo_RETS}. In each training iteration, a test subset is randomly selected from $\bm{D}_{train}^{0}$, which contains samples of all base classes. The test subset set is used as test data in a pseudo-incremental session containing test samples of pseudo-old and pseudo-new classes. Multiple training subsets are randomly selected from $\bm{D}_{train}^{0}$ and then used to obtain the prototypes of pseudo-new classes. Meanwhile, the prototypes in $\bm{P}_{0}$ whose labels are different from those of the pseudo-new classes, are used as the pseudo-base prototypes. Then, the initial pseudo-incremental prototypes are obtained by merging the pseudo-new and pseudo-base prototypes. The refined pseudo-incremental prototypes are obtained using the DRPM. After that, the predictions of the test subset can be made using the EE, RM, and refined pseudo-incremental prototypes. Finally, the cross-entropy losses of multiple episodes are calculated, and the model is optimized via the algorithm of stochastic gradient descent (SGD) \cite{lei2021learning}.

\begin{figure*}[ht]
  \centering
  \includegraphics[scale=0.24]{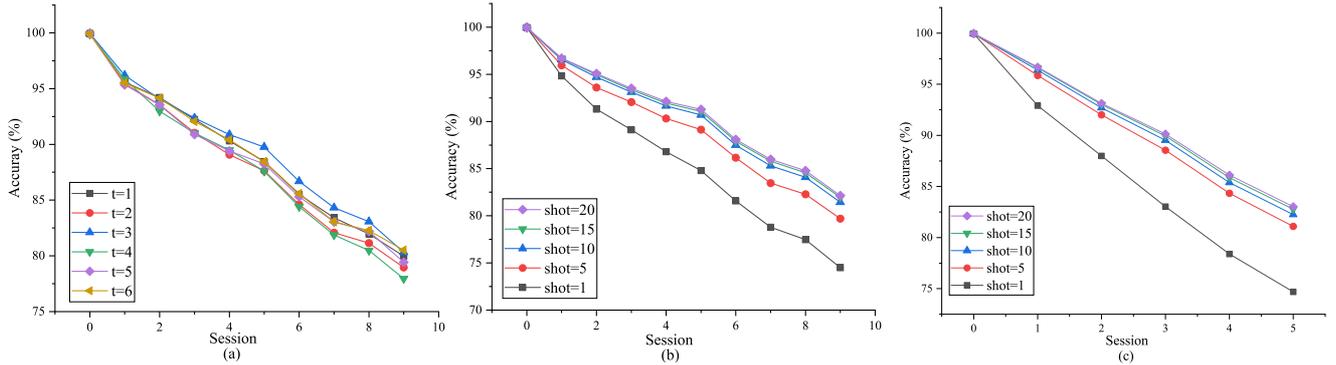}
  \caption{\footnotesize Performance analysis of our method with different settings on the Nsynth-100. (a) The impact of the episode number in Algorithm \ref{algo_RETS}. (b) Performance of our method in the setting of 5-way with different numbers of shots. (c) Performance of our method in the setting of 9-way with different numbers of shots.}
  \label{fig_further_results}
\end{figure*}

\subsection{Dynamic relation projection module}
\label{ssec:DRP}

The initial prototypes $\bm{P}_{init} \in \mathbb{R} ^{N_{init}\times d}$ in each incremental session are constructed by merging the new prototypes and the prototypes of previous session $\bm{P}_{pre}\in \mathbb{R} ^{N_{pre}\times d}$. To enhance the discriminability of the initial prototypes, the DRPM employs a set of relation weights $\bm{M} \in \mathbb{R} ^{N_{init}\times N_{pre}}$ to refine the initial prototypes in each incremental session. $N_{init}$ and $N_{pre}$ are the number of audio classes in the current and previous sessions, respectively. The relation weights are obtained by
\begin{equation}
  \bm{M}=f_{1}\left ( \bm{P}_{init} \right ) \left ( f_{2} \left (   \bm{P}_{pre}  \right ) \right ) ^{T}
\end{equation}
where $f_{1}$ and $f_{2}$ represent two convolutional blocks for transforming the prototypes to a shared latent space. Each block contains a $1\times 1$ convolutional layer, a batch normalization layer, and a ReLU layer. The value in $\bm{M}$ represents the similarity between different prototypes, so it can be used to weight the prototypes to achieve refinement of the prototypes. Therefore, the refined prototypes $\bm{P}_{re} \in \mathbb{R} ^{N_{init}\times d}$ is obtained by $ \bm{P}_{re}=\bm{M}\bm{P}_{pre} $.

\section{Experiments}
\label{sec:experiments}

\subsection{Experimental setup}
\label{ssec:datasets}

The Nsynth-100 and FSC-89 datasets\footnote{https://github.com/chester-w-xie/FCAC\_datasets}  are derived from the public audio corpora of Nsynth \cite{engel2017neural} and FSD-MIX-CLIPS \cite{wang2021calls}, respectively. Statistics of the experimental datasets are shown in Table. \ref{tb_datasets}. The baseline methods of Finetune \cite{zhou2021pycil}, ICaRL \cite{rebuffi2017icarl}, and DFSL \cite{gidaris2018dynamic} are re-implemented with open source codes according to the problem setup of FCAC. For a fair comparison, the backbone of the ResNet18 \cite{he2016deep} is used as the EE for all methods. Meanwhile, FBFs is used in all methods. The RM used in the proposed method consists of three $1\times1$ convolutional layers with a batch normalization, a ReLU layer, and a dropout layer between every two convolutional layers. Accuracy in each session $\mathcal{A} _l$, average accuracy over all sessions $AA=\frac{\sum_{l=0}^{L}\mathcal{A} _l}{L} $, and performance dropping rate $PD= \mathcal{A} _0-\mathcal{A} _L$ \cite{zhang2021few} are used as performance metrics. $\mathcal{A} _0$ is the accuracy in the base session, and $\mathcal{A} _L$ is the accuracy in the last incremental session. All results are obtained by averaging the outputs of 100 trials.
\begin{table}[h]
  \caption{Statistics of the experimental datasets. NBS and NNS denotes the number of training, validation and testing samples per base and new class, respectively.}
  \label{tb_datasets}
  \centering
  \renewcommand{\arraystretch}{1.5}
  \resizebox{.47\textwidth}{!}{
  \begin{tabular}{lcc}
    \toprule[1.5pt]
    & NSynth-100             & FSC-89                                       \\ \hline
  Type of sound                                                   & Musical instruments                           & Sound events                                  \\ \hline
  No. of classes             & 55 (base), 45 (new)  & 59 (base), 30 (new) \\ \hline
  NBS  & 200, 100, 100                  & 800, 200, 200                              \\ \hline
  NNS & 100 / none / 100                              & 500 / none / 200                             \\ \hline
  Duration per sample                                          &  4 seconds                              & 1 second                              \\ \hline
  Sampling frequency                        & 16 kHz                                 & 44.1 kHz                              \\ \toprule[1.5pt]
  \end{tabular}
  }
\end{table}
\vspace{-1.5em}

  \subsection{Results and discussions}
  \label{ssec:re_dis}

  \begin{table*}[h]
    
    \caption{Results obtained by different methods on the Nsynth-100.}
    \label{tb_Comparison_resullts_nsynth100}
    \centering
    \footnotesize
    \renewcommand{\arraystretch}{1.5}
    \begin{tabular}{lcccccccccccc}
      \toprule[1.5pt]
    \multirow{2}{*}{Method} & \multicolumn{10}{c}{Accuracy in each session (\%) $\uparrow$}                              & \multirow{2}{*}{AA $\uparrow$} & \multirow{2}{*}{PD $\downarrow$} \\ \cline{2-11}
           & 0  & 1  & 2 & 3  & 4  & 5  & 6   & 7 & 8 & 9  &                     &                     \\ \hline
    Finetune                & 99.96 & 84.73 & 76.92 & 71.06 & 63.18 & 40.15 & 47.99 & 38.33 & 38.01 & 37.86 & 59.82               & 62.10                \\ \hline
    ICaRL                   & \textbf{99.98} & 93.30 & 88.88 & 84.82 & 79.57 & 67.65 & 65.92 & 59.65 & 54.62 & 51.01 & 74.54               & 48.97               \\ \hline
    DFSL                    & 99.91 & 95.36 & 91.51 & 88.02 & 84.30 & 81.97 & 79.08 & 77.12 & 76.55 & 75.14 & 84.89               & 24.77               \\ \hline
    Proposed                    & 99.96 & \textbf{95.95} & \textbf{93.60} & \textbf{92.06} & \textbf{90.32} & \textbf{89.14} & \textbf{86.16} & \textbf{83.47} & \textbf{82.28} & \textbf{79.69} & \textbf{89.26}               & \textbf{20.27}               \\ \toprule[1.5pt]
    \end{tabular}
    \vspace{-2.2em}
    \end{table*}

  We first perform comparison experiments on the two audio datasets. The results obtained by different methods on the Nsynth-100 and the FSC-89 are shown in Table \ref{tb_Comparison_resullts_nsynth100} and Table \ref{tb_Comparison_resullts_fsc89}, respectively. As shown in Table \ref{tb_Comparison_resullts_nsynth100}, the accuracies achieved by our method are higher than that of three baseline methods in all nine incremental sessions. The AA obtained by our method is 89.26\%, which is 29.44\%, 14.72\%, and 4.37\% higher than that of the Finetune, ICaRL, and DFSL methods, respectively. In addition, the PD obtained by our method is 20.27\%, which is 41.83\%, 28.70\%, and 4.5\% lower than that of the Finetune, ICaRL, and DFSL methods, respectively. Because the Finetune method uses a few training samples of new classes to retrain the model, the model overfits the new classes and catastrophically forgets the learned classes. As a result, the Finetune method performs worse than other methods. With data retention and knowledge distillation strategies, the ICaRL method obtains higher classification accuracies than the Finetune method. However, since the ICaRL method also employs the retraining strategy to expand the classifier, the model cannot effectively remember the learned classes' knowledge when new classes appear continuously. Therefore, with the increase of new classes, its performance worsens. To adapt the DFSL method for the FCAC problem, we reuse an ABWG trained in the base session to generate new classification weights in each incremental session. The parameters of the ABWG were frozen in all incremental sessions for avoiding interference caused by the retraining strategy. However, without a powerful EE and the refinement of the classification weights, the model's classification performance can hardly be improved. In contrast, the proposed method employs the RETS to train the model to improve its generalization ability, which enables EE to generate the initial prototypes with strong discriminability. Moreover, the proposed method employs the DRPM to refine the prototypes in incremental sessions to further enhance their discriminability .

  \begin{table}[h]
    \caption{Results obtained by different methods on the FSC-89.}
    \label{tb_Comparison_resullts_fsc89}
    \centering
    \renewcommand{\arraystretch}{1.5}
    \resizebox{.49\textwidth}{!}{
    \begin{tabular}{lccccccccc}
      \toprule[1.5pt]
    \multirow{2}{*}{Method} & \multicolumn{7}{c}{Accuracy in each session (\%)$\uparrow$}       & \multirow{2}{*}{AA$\uparrow$} & \multirow{2}{*}{PD$\downarrow$} \\ \cline{2-8}
                            & 0     & 1     & 2     & 3     & 4     & 5     & 6     &                     &                     \\ \hline
    Finetune               & 42.28 & 31.67 & 29.11 & 24.63 & 21.74 & 20.17 & 16.70 & 26.61               & 25.58                \\ \hline
    ICaRL                  & \textbf{42.43} & 36.25 & 33.00 & 26.60 & 26.63 & 23.76 & 20.71 & 29.91               & 21.72               \\ \hline
    DFSL                     & 42.36 & 35.23 & 32.72 & 31.21 & 29.79 & 28.51 & 27.40 & 32.46               & 14.96               \\ \hline
    Proposed                   & 42.04 & \textbf{39.95} & \textbf{37.01} & \textbf{34.68} & \textbf{32.97} & \textbf{31.45} & \textbf{30.09} & \textbf{35.46}               & \textbf{11.95}               \\ \toprule[1.5pt]
    \end{tabular}
    }
    \vspace{-1.0em}
    \end{table}

  Since the FSC-89 is more complex than the Nsynth-100 in terms of background noise, inter-class confusion, and intra-class inconsistency, all methods obtain lower accuracy on the FSC-89 than on the Nsynth-100, as shown in Table \ref{tb_Comparison_resullts_nsynth100} and Table \ref{tb_Comparison_resullts_fsc89}. However, the advantages of the proposed method over the baseline methods still hold for the FSC-89 in terms of accuracy, AA, and PD. 

  To show the performance of the proposed method with different experimental settings, we conducted three performance analysis experiments on the Nsynth-100. Fig. \ref{fig_further_results} (a) shows the impact of different numbers of episode $t$ on the performance of the proposed method. The results show that the proposed method obtains the best overall performance when $t=3$. Fig. \ref{fig_further_results} (b) and Fig. \ref{fig_further_results} (c) present the accuracy obtained by the proposed method with different numbers of shots. By comparing Fig. \ref{fig_further_results} (b) and Fig. \ref{fig_further_results} (c), it can be found that more shots lead to better performance. Moreover, when the model needs to recognize more new audio classes within a session, its classification performance decreases more severely, which is the reason why the curve in Fig. \ref{fig_further_results} (c) is sharper than that in Fig. \ref{fig_further_results} (b).

  \section{Conclusions}
\label{sec:conclusions}

This paper addresses a new problem of audio classification termed few-shot classes-incremental audio classification. A method to expand the model with adaptively refined prototypes is proposed for the problem. We also adapt the methods of finetune, ICaRL, and DFSL to the FCAC problem and use them as the baseline methods. Experimental results on the NSynth-100 and the FSC-89 datasets indicate that our method achieves state-of-the-art performance. This work can be a benchmark for future studies of the FCAC problem. Our future works include designing a more effective prototype refinement strategy and evaluating the methods with more comprehensive experiments.


\bibliographystyle{IEEEtran}
\bibliography{mybib}

\end{document}